 \documentstyle[preprint,aps,prl]{revtex}
\begin{document}
\draft
\preprint{SNUTP  96-094}
\title{Color-octet mechanism in the inclusive 
$D-$wave \\ charmonium productions  in $B$  decays}
\author{ Pyungwon Ko$^{1,}$\footnote{ pko@phyb.snu.ac.kr}}
\address{$^{1}$Department of Physics, Hong-Ik University,
Seoul 121-791, Korea }
\author{Jungil Lee$^{2,}$\footnote{jungil@phyb.snu.ac.kr}
    and H.S. Song$^{2,}$\footnote{hssong@physs.snu.ac.kr} }
\address{
$^{2}$Department of physics and Center for Theoretical Physics
\\ Seoul National University,
Seoul 151-742, Korea  }
\date{\today}
\maketitle
\begin{abstract}
The inclusive $D-$wave charmonium production rates  in $B$ decays
are considered in the Bodwin-Braaten-Lepage (BBL) approach.
We find that the color-octet subprocesses
$B\to c\overline{c}({^3S_1^{(8)}}{\rm or} ~{^3P_J^{(8)}})+s,d$, followed by
the transition $c\overline{c}({^3S_1^{(8)}},{\rm or}~{^3P_J^{(8)}})\to~^3D_J$,
strongly dominate over any other subprocess, due to the large Wilson 
coefficient for the $ \Delta B  = 1$ effective lagrangian.  
Assuming that the numerical values of  the matrix elements 
$\langle0|{\cal O}^{^3D_J}(^3S_1^{(8)})|0\rangle$
and
$\langle0|{\cal O}^{^3D_J}(^3P_J^{(8)})|0\rangle$ are the same order of 
magnitudes with the $\langle0|{\cal O}^{\psi^{'}}(^3S_1^{(8)})|0\rangle$,
we find that the ${^{1,3}D_{2}}$ can be observed at future B-factories.
\end{abstract}
\newpage
%
{\bf 1.}
Estimates of $S-$ and $P-$wave charmonium production rates in $B$ decays 
were one of the earliest applications 
of the heavy quarkonium physics in the framework of perturbative QCD (PQCD)
\cite{kuhn}.  In the case of $P-$wave charmonia, it had long been known that 
their decay and production rates in the color-singlet model have
infrared singularities, which signaled the failure of the factorization 
procedure in terms of a single wavefunction of the $c\overline{c}$ 
pair in the color-singlet $P-$wave state.
Recent development in Nonrelativistic QCD (NRQCD) as an effective field 
theory of QCD \cite{NRQCD} helped such inconsistency  be resolved  
including higher Fock states of the $P-$wave charmonia in a 
systematic manner in NRQCD~\cite{pdecay}~\cite{bodwinB}.

On the contrary, the $S-$wave charmonium productions in $B$ decays 
have no such infrared divergence
problems.   However, the prediction based on the
color-singlet model falls short of the data by a factor of $\sim 3$. 
This can be partly overcome by including the color-octet mechanism, since the
relevant Wilson coefficient is a factor of $\sim 30$ larger than that of the
color-singlet case \cite{jungilb}.  
Also the polarization of $J/\psi$ produced in $B$
decays depend sensitively on the color-octet contributions  \cite{flemingB}.

In this work, we extend previous works~\cite{bodwinB}~\cite{jungilb}
\cite{flemingB} to the case of the $D-$wave 
charmonium productions in $B$ decays.  In various potential models for the 
charmonium family, $D-$wave charmonium states are predicted in the mass 
range of $3.81-3.84$ GeV \cite{rosner}.  Although the masses are above the 
$D\bar{D}$ threshold, they are below the $D \bar{D}^{*}$ threshold. 
Also,  two states
${^3D_2} (J^{PC} = 2^{--})$ and ${^1D_2} (J^{PC} = 
2^{-+})$ are forbidden to decay into $D\bar{D}$ because of parity 
conservation.  So these two states are predicted to have narrow decay widths, 
and we can easily tag these two states through dominant decay channels 
(to be discussed below).
Therefore,  $D-$wave charmonium states (at least for $J=2$) is certainly 
accessible at 
$B$-factories where one expects to have $\sim 10^8$ $B$ decays per year.
Thus $B$ factories can serve as a place for studying
new charmonium states below/above the $D \bar{D}$ threshold, including 
some hypothetical states like hybrids or $\chi_{c_J}(2P)$ state \cite{ko2p}, 
in addition to its main role for measurements of CP asymmetries in $B$ 
decays.

{\bf 2.}  The $\Delta B  = 1$ effective Hamiltonian for 
$b \rightarrow c \overline{c} q$ (
with $q = d,s$) is written as~\cite{bodwinB}
\begin{eqnarray}
H_{eff} & = & {G_{F} \over \sqrt{2}}~V_{cb} V_{cq}^{*}~
\left[ {{2 C_{+} - C_{-}}
\over 3}~\overline{c} \gamma_{\mu} (1 -\gamma_{5})c~\overline{q} 
\gamma^{\mu} (1 -
\gamma_{5}) b ~ \right.
\nonumber    \\
 & + & \left. (C_{+} + C_{-})~\overline{c} \gamma_{\mu} (1 - \gamma_{5})
T^{a} c ~\overline{q}\gamma^{\mu} (1 - \gamma_{5}) T^{a} b \right],
\label{eq:H_eff}
\end{eqnarray}
where $C_{\pm}$'s are the Wilson coefficients at the scale $\mu \approx
M_b$. We have neglected penguin operators, since their Wilson coefficients
are small  and thus they are irrelevant to our case.
To the leading order in $\alpha_{s}(M_{b})$ and to all orders in $\alpha_{s}
(M_{b})~{\rm ln}(M_{W}/M_{b})$,  the above Wilson coefficients 
are~\cite{bodwinB}
\begin{equation}
C_{+} (M_{b}) \approx 0.87,~~~~~ C_{-} (M_{b}) \approx 1.34.
\end{equation}

At the parton level, the $B$ decay into ${^{2S+1}D}_J$ charmonium 
state occurs through 
\begin{eqnarray}
b&\to&(c\overline{c}){^{2S+1}L}_J^{(1,8)}+s,d \to~{^{2S+1}D}_J+X.
\end{eqnarray}
Let us first estimate the velocity scaling of the various subprocesses. From
the effective Hamiltonian given in Eq.~(\ref{eq:H_eff}), one can get the 
amplitude for $b\to(c\overline{c}){^{2S+1}L}_J^{(1,8)}+s,d $ by the Taylor 
expansion of the amplitude with respect to the relative velocity ($v$)
of  the heavy quark inside the heavy quarkonium and projection over the 
specific partial wave state of $(c\overline{c})$.
Expanding  the physical ${^{2S+1}D_J}$ state in terms of 
higher Fock states as following, 
\begin{eqnarray}
|^3D_J\rangle
&=& 
 O(1  )|c\overline{c}(^3D_J^{(1)})\rangle
+O(v^1)|c\overline{c}(^3P_J^{(8)})g\rangle
+O(v^1)|c\overline{c}(^3F_J^{(8)})g\rangle
\nonumber\\
&+&
 O(v^2)|c\overline{c}(^3D_J^{(8)})gg\rangle
+O(v^2)|c\overline{c}(^3S_1^{(1,8)})gg\rangle
+O(v^2)|c\overline{c}(^1D_2^{(8)})g\rangle
+\ldots,
\\
|^1D_2\rangle
&=&
 O(1  )|c\overline{c}(^1D_2^{(1)})\rangle
+O(v^1)|c\overline{c}(^1P_1^{(8)})g\rangle
+O(v^1)|c\overline{c}(^1F_3^{(8)})g\rangle
\nonumber\\
&+&
 O(v^2)|c\overline{c}(^1D_2^{(8)})gg\rangle
+O(v^2)|c\overline{c}(^1S_0^{(1,8)})gg\rangle
+O(v^2)|c\overline{c}(^3D_J^{(8)})g\rangle
+\ldots,
\end{eqnarray}
we can find the velocity scaling of the soft process.  Note that the 
chromoelectric E1 and the chromomagnetic M1 transitions accompany with
the extra factors of $v$ and $v^2$, respectively, in the amplitude for 
the soft transition, $(c\overline{c}) ({^{2S+1}L_{J}}^{1,8}) \rightarrow
{^{2S+1}D_J}$. 
In Table~\ref{table:scaling}, we present the velocity scalings of the 
amplitude in NRQCD for  various channels.
According to Table~\ref{table:scaling}, 
the following  processes are leading order in velocity scaling :
\begin{eqnarray}
\left.
\begin{array}{cclcl}
b&\to&(c\overline{c})~^{3}D_J^{(1)}&  +&s,d\\
b&\to&(c\overline{c})~^{3}P_J^{(8)}&  +&s,d\\
b&\to&(c\overline{c})~^{3}S_1^{(1,8)}&+&s,d
\end{array}
\right\}&\to&~^3D_J+X,
\\
\left.
\begin{array}{cclcl}
b&\to&(c\overline{c})~^{1}D_2^{(1)}  &+&s,d\\
b&\to&(c\overline{c})~^{1}S_0^{(1,8)}&+&s,d
\end{array}
\right\}&\to&~^1D_2+X.
\end{eqnarray}
Note that in the hard process amplitude for 
$B\to(c\overline{c})^{2S+1}L_J+s,d$, the spectroscopic states
$(c\overline{c})^1P_1^{(1,8)}$ do not contribute in the leading order.
Since the Wilson coefficient for the 
color-octet subprocess is about 30 times larger than that of the 
color-singlet subprocess in the (amplitude)$^2$ level\cite{jungilb},
we find that the following subprocesses (which are $O(v^2)$ in the 
amplitude level) dominate over all the others : 
\begin{eqnarray}
\left.
\begin{array}{cclcl}
b&\to&(c\overline{c})^{3}P_J^{(8)}&+&s,d\\
b&\to&(c\overline{c})^{3}S_1^{(8)}&+&s,d
\end{array}
\right\}
&\to&~^3D_J+X,
\\
\begin{array}{cclcl}
b&\to&(c\overline{c})^{1}S_0^{(8)}&+&s,d
\end{array}
\hskip4mm
&\to&~^1D_2+X,
\end{eqnarray}
Using the results in Ref.~\cite{jungilb},
we immediately obtain the inclusive $^{2S+1}D_J$ charmonium 
production rate in $B$ decay as
\begin{eqnarray}
\Gamma(B\to~^{3}D_J+X)
&\simeq&
\Gamma(B\to(c\overline{c})^{3}S_1^{(8)}+s,d\to~^3D_J+X)
\nonumber\\
&+&
\Gamma(B\to(c\overline{c})^{3}P_J^{(8)}+s,d\to~^3D_J+X)
\nonumber\\
&=&
~(C_{+} + C_{-})^{2}
~\left( 1 + {8 M_{c}^{2} \over M_{b}^2} \right)
~\hat{\Gamma}_0
\nonumber\\
&\times&
\left[
\frac{\langle 0 | O^{^3D_J} (^{3}S_{1}^{(8)}) | 0 \rangle}{2 M_{c}^2}
+
\frac{3\langle 0 | O^{^3D_J} (^{3}P_{0}^{(8)}) | 0 \rangle}{M_{c}^4}
\right],
\\
\Gamma(B\to~^{1}D_2+X)
&\simeq&
\Gamma(B\to(c\overline{c})^{1}S_0^{(8)}+s,d\to~^1D_2+X)
\nonumber\\
&=&
~(C_{+} + C_{-})^{2}
~\hat{\Gamma}_0~
~\frac{3\langle 0 | O^{^1D_2} (^{1}S_{0}^{(8)}) | 0 \rangle}{2 M_{c}^2}
,
\end{eqnarray}
where
\begin{equation}
\hat{\Gamma}_{0} \equiv |V_{cb}|^{2} \left( {G_{F}^{2} \over 144 \pi}
\right) M_{b}^{3} M_{c} \left( 1 - {4 M_{c}^{2} \over M_{b}^2} \right)^{2}.
\end{equation}
In order to evaluate the branching ratios numerically, we need to know three
nonperturbative matrix elements, $\langle 0 | O^{^3D_J} (^{3}S_{1}^{(8)}) 
| 0 \rangle$, 
$\langle 0 | O^{^3D_J} (^{3}P_{0}^{(8)}) | 0 \rangle$ and
$\langle 0 | O^{^1D_2} (^{1}S_{0}^{(8)}) | 0 \rangle$.

In connection  with this, we note that Qiao {\it et. al.} \cite{qiao} 
have recently shown that the color-octet production processes
$Z^0\to {^3D_J} (c\overline{c})q\overline{q}$
have distinctively large branching ratios and it is
of the same order as that of the $J/\psi$ production in $Z^0$ decays, 
assuming  
\begin{eqnarray}
\langle0|{\cal O}^{^3D_J}(^3S_1^{(8)})|0\rangle
&=&\frac{(2J+1)}{5} \langle0|{\cal O}^{\psi^\prime}(^3S_1^{(8)})|0\rangle
\nonumber   \\
&=&\frac{(2J+1)}{5}\times 4.6\times10^{-3}~{\rm GeV}^3, 
~~~~{\rm for}~J=1,2,3.
\end{eqnarray}
Similarly, we can estimate the numerical value of the matrix element
$\langle0|{\cal O}^{^3D_J}(^3P_0^{(8)})|0\rangle$ as
\begin{eqnarray}
\langle0|{\cal O}^{^3D_J}(^3P_0^{(8)})|0\rangle
&=&\frac{(2J+1)}{5} \langle0|{\cal O}^{\psi^\prime}(^3P_0^{(8)})|0\rangle
\nonumber\\
&=&\frac{(2J+1)}{5}\times 1.3\times10^{-2}~{\rm GeV}^5
~~~~{\rm for}~J=1,2,3,
\end{eqnarray}
using the numerical value which was maximally allowed in Ref.~\cite{pcho2}. 
This value is about an order of magnitude larger than that deduced from the 
velocity scaling.  For the case of $^1D_2$, we use the numerical value
of $\langle 0 | O^{^1D_2} (^{1}S_{0}^{(8)}) | 0 \rangle$ by
imposing the approximate heavy quark spin symmetry as 
\begin{eqnarray}
\langle 0 | O^{^1D_2} (^{1}S_{0}^{(8)}) | 0 \rangle 
&=&\langle0|O^{^3D_2} (^{3}S_{1}^{(8)})|0\rangle.
\end{eqnarray}
With the numerical values as
\begin{eqnarray}
M_{b} = 5.3 ~{\rm GeV}, && M_c = 1.5~{\rm GeV},\\
\alpha_s(2M_c)&=&0.253,\\
\langle 0 | O^{^3D_2} (^{3}S_{1}^{(8)}) | 0 \rangle
&=&4.6\times10^{-3}~{\rm GeV}^3,
\end{eqnarray}
we estimate the branching ratios as
\begin{eqnarray}
B(B\to~^3D_J+X)&=&\frac{2J+1}{5}\times
\left\{
\begin{array}{ccl}
5.2\%&{\rm when}&
\langle 0 | O^{^3D_2} (^{3}P_{0}^{(8)}) | 0 \rangle
=1.3\times10^{-2}{\rm GeV}^5\\
1.1\%&{\rm when}&
\langle 0 | O^{^3D_2} (^{3}P_{0}^{(8)}) | 0 \rangle
=1.3\times10^{-3}{\rm GeV}^5
\end{array}\right.,
\\
B(B\to~^1D_2+X)&=&0.67\%.
\end{eqnarray}

Before closing this subsection, we have to mention on theoretical
uncertainties in our estimates for $B(B \rightarrow D + X)$. The first 
comes from the assumptions, (13)--(15), which originate from our ignorance 
of nonperturbative physics in QCD.  Up to now, there are no known processes 
which depend on these three parameters. So we relied upon the naive guess 
based on  the velocity scaling rule and heavy quark spin symmetry. Another
intrinsic uncertainty comes from the approach adopted here and  in 
Ref.~\cite{bodwinB}. Namely, the parton picture adopted here 
for the case of $D-$wave charmonium productions in 
$B$ decays may not be so good as in the case of the $S-$wave charmonium 
productions in Ref.~\cite{jungilb} 
because the phase space available in the former is much less than 
the latter.  
There are also uncertainties coming from the quark masses $M_b$, $M_c$ and 
the relation between $M_c$ and the mass of the $D-$wave charmonium state. 
In order to minimize these last uncertainties, 
we have  used the prescriptions made in Ref.~\cite{bodwinB}.  Despite of 
all of these uncertainties,  we still anticipate  our results to be 
correct estimates of {\it an order of magnitude } for the branching ratios 
for the decays $B \rightarrow D + X$.   

{\bf 3. }
In order to identify the desired $D-$wave charmonium state produced in 
$B$ decays, one has to know its decay channels. As mentioned earlier, 
$J=2$ states (${^{2S+1}D_2}$) are predicted to have narrow decay widths
of order of $300-400$ keV.  So these states are easier to reconstruct in the
experiments, and hereafter we consider only $J=2$ $D-$wave states, and  
try to find out which mode is best to identify
the $J=2$ $D-$wave charmonium state in $B$ decays.  

First consider the spin-triplet $D-$wave state ($J^{PC} = 2^{--}$).
The main decay modes of this state are the E1 transitions into 
$\chi_{c_{J=1,2}}
(1P)$ states, the hadronic transition into $J/\psi + \pi \pi$, and the
decay into light hadrons via $c\overline{c}$ annihilation into three gluons.
The decay widths for each process is rather model dependent, but it is 
enough to quote some nominal value for each mode, in order to have some 
idea about the branching ratio for each channel.  
We adopt the numbers given in Ref.~\cite{qiao} :
\begin{eqnarray}
B ({^3D_2} \rightarrow \chi_{c1} + \gamma ) = 0.64,
\nonumber   \\
B ({^3D_2} \rightarrow \chi_{c2} + \gamma ) = 0.15,
\\
B ( {^3D_2} \rightarrow J/\psi + \pi^+ \pi^- ) = 0.12.
\nonumber  
\end{eqnarray}
Combining the branching ratios for $\chi_{c_{J=1,2}} (1P) \rightarrow 
J/\psi + \gamma$ and $J/\psi \rightarrow \mu^{+} \mu^{-}$, we are led to 
\begin{eqnarray}
B ( {^3D_2} \rightarrow J/\psi + \pi^+ \pi^- ) \times 
B ( J/\psi \rightarrow \mu^{+} \mu^{-} ) &=& 7 \times 10^{-3},
\nonumber    \\
B ( {^3D_2} \rightarrow  \chi_{c1} + \gamma ) \times
B ( \chi_{c1} \rightarrow J/\psi + \gamma ) \times
B ( J/\psi \rightarrow \mu^{+} \mu^{-} ) &=& 10.4 \times 10^{-3},
\\
B ( {^3D_2} \rightarrow  \chi_{c2} + \gamma ) \times
B ( \chi_{c2} \rightarrow J/\psi + \gamma ) \times
B ( J/\psi \rightarrow \mu^{+} \mu^{-} ) &=& 1.2 \times 10^{-3}.
\nonumber 
\end{eqnarray}
The last decay channel through the intermediate $\chi_{c2} (1P)$ state 
is negligible. For the radiative decay channels, one needs a good 
photon detector. Otherwise the first option through ${^3D_2} \rightarrow 
J/\psi + \pi^+ \pi^-$ would be the best to reconstruct ${^3D_2}$ state 
at $B$ factories.  
Combining the branching ratios for the cascade decays, we get 
\begin{eqnarray}
&&B(B\to~^3D_2\to J/\psi(\to\mu^{+} \mu^{-}) + \pi^+\pi^-)
\nonumber\\
&&\hskip1cm\sim
\left\{
\begin{array}{ccl}
3.7\times10^{-4}&{\rm when}&
~\langle 0 | O^{^3D_2} (^{3}P_{0}^{(8)}) | 0 \rangle
=1.3\times10^{-2}{\rm GeV}^5,\\
7.5\times10^{-5}&{\rm when}&
~\langle 0 | O^{^3D_2} (^{3}P_{0}^{(8)}) | 0 \rangle
=1.3\times10^{-3}{\rm GeV}^5.
\end{array}\right.
\end{eqnarray}
Comparing these numbers with those for $B \rightarrow J/\psi ({\rm or }~
\psi{'}) + X$,
\begin{eqnarray}
B(B\to J/\psi+X\to \mu^+\mu^-+X)&=&4.7\times10^{-4}, \\
B(B\to \psi^\prime+X\to \mu^+\mu^-+X)&=&2.0\times10^{-4},
\end{eqnarray}
we find that the $D-$wave charmonium productions in $B$ decays are not so
rare, and can be within the reach of B-factories.
There would be  several tens to a few hundreds' ${^{3}D_2}$'s in $10^6 ~B$ 
meson decays into $B\to ~^3D_2\to J/\psi \pi^+\pi^-$ with $J/\psi \rightarrow
\mu^+ \mu^-$, depending on the numerical values of 
$\langle 0 | O^{^3D_2} (^{3}S_{1}^{(8)}) | 0 \rangle$ and 
$\langle 0 | O^{^3D_2} (^{3}P_{0}^{(8)}) | 0 \rangle$.

Next consider the spin-singlet $D-$wave state ($J^{PC} = 2^{-+}$).
One possible decay channel for this state is  \cite{wise}  
\begin{eqnarray}
B ( {^1D_2} \rightarrow h_{c} ({^1P_1}) + \gamma )  & = &  0.80 
\nonumber   \\
B ( h_{c} ({^1P_1}) \rightarrow J/\psi + \pi^{0} )  & = &  0.005
\\
B ( J/\psi \rightarrow \mu^{+} \mu^{-} )            & = & 0.06.
\nonumber
\end{eqnarray}
Here one has to tag three photons, one from the first chain, and the
other two from the $\pi^0$ decay.  The product of three branching ratios 
in this channel is $2.4 \times 10^{-4}$, so that we expect    
\begin{eqnarray}
B(B\to~^3D_2\to \gamma+h_c(\to \pi^0+J/\psi(\to\mu^{+} \mu^{-})))
&\sim& 1.6\times10^{-6},
\end{eqnarray}
corresponding to $\sim 160$ such cascade events in $10^{8} B$ decays.


{\bf 4}. 
In conclusion, we have considered the $D-$wave charmonium productions
in $B$ decays. The dominant contributions come from the color-octet
$(c\overline{c}){^3S_1^{(8)}}$
and $(c\overline{c})^3P_J^{(8)}$ states for $^3D_J$ and 
$(c\overline{c}){^1S_0^{(8)}}$ state for $^1D_2$ production, respectively, 
because of the large 
Wilson coefficient, $(C_{+} + C_{-})$ in Eq.~(\ref{eq:H_eff}).  Assuming 
$\langle 0 | O^{^3D_J} ({^3S_1}^{(8)}) | 0 \rangle = \langle 0 | O^{\psi^{'}} 
({^3S_1}^{(8)}) | 0 \rangle \times (2J+1) / 5 $ and 
$\langle 0 | O^{^3D_J} ({^3P_0}^{(8)}) | 0 \rangle = \langle 0 | O^{\psi^{'}}
({^3P_0}^{(8)}) | 0 \rangle \times (2J+1) / 5 $, we obtain 
$B ( B \rightarrow {^3D_2} + X) \approx 1.1 \% - 5.1 \%$.  
Assuming $B( {^3D_2} 
\rightarrow J/\psi + \pi^+ \pi^- ) \approx 10 \%$ and using $B (
J/\psi \rightarrow 
\mu^{+} \mu^{-}) = 0.06$,  one can find 75 to 370 ${^3D_2}$'s  
in the $10^6~B$ decays into $B \rightarrow {^3D_2} + X$ followed by 
$ {^3D_2} \rightarrow J/\psi + \pi^+  \pi^-$ and $J/\psi \rightarrow
l^{+} l^{-}$ with $l = \mu$. This is quite accessible in 
$B$-factories
where $10^{8}$ $B$ meson decays will be accumulated per year 
\footnote{ Here, we have neglected the detector efficiencies for 
simplicity. Typically, the detector efficiency for $J/\psi \rightarrow 
\mu^+ \mu^-$ is less than $\sim 50 \%$, and that for $\pi^0$ is even less.
So our numbers of events should be reduced by such efficiencies. Also, one
can use $J/\psi \rightarrow e^+ e^-$ for $J/\psi$ tagging, which would
help double the number of events from our estimates.}.  
Since the $D-$wave charmonium states are not easy  to find at other colliders,
it is recommended to search for these states as well as others
such as $\chi_{c_{J}}(2P)$ and hybrid charmonia from $B$ decays.  
Finding such states at the level predicted in this work will be 
another test of the factorization based on the NRQCD, including the 
color-octet mechanism for the heavy quarkonium productions.
Also it will deepen our understanding of the heavy quarkonium spectroscopy 
based on the color-singlet pair and exotic hybrid states.

\acknowledgements
We are grateful to Prof. Youngjoon Kwon for useful comments and suggestions
on the experimental aspects in B-factories.
This work was supported in part by KOSEF through CTP at Seoul National 
University,
and in part by the Basic Science Research
Program, Ministry of Education,  Project No. BSRI--96--2418.
P.K. is also supported in part by NON DIRECTED RESEARCH FUND 1995, 
Korea Research  Foundations.
J.L. is supported in part by  Korea Research Foundations 
through Post Doc. Program.

\begin{table}
\begin{tabular}{ccccc}
 subprocess           & $L$ & $E$ & $M$  & total ($v^{L+E+2M}$)
\\   \tableline
$b\to~(c\overline{c}) ^{3}D_J^{(1  )}+s,d\to~^{3}D_J+X$&$2$&$0$&$0$&$v^2$\\
$b\to~(c\overline{c}) ^{3}D_J^{(  8)}+s,d\to~^{3}D_J+X$&$2$&$2$&$0$&$v^4$\\
$b\to~(c\overline{c}) ^{1}D_2^{(8  )}+s,d\to~^{3}D_J+X$&$2$&$0$&$1$&$v^4$\\
$b\to~(c\overline{c}) ^{1}P_1^{(1,8)}+s,d\to~^{3}D_J+X$&$1$&$1$&$1$&$v^4$\\
$b\to~(c\overline{c}) ^{3}P_J^{(  8)}+s,d\to~^{3}D_J+X$&$1$&$1$&$0$&$v^2$\\
$b\to~(c\overline{c}) ^{1}S_0^{(1,8)}+s,d\to~^{3}D_J+X$&$0$&$2$&$1$&$v^4$\\
$b\to~(c\overline{c}) ^{3}S_1^{(1,8)}+s,d\to~^{3}D_J+X$&$0$&$2$&$0$&$v^2$\\
\hline
$b\to~(c\overline{c}) ^{1}D_2^{(1  )}+s,d\to~^{1}D_2+X$&$2$&$0$&$0$&$v^2$\\
$b\to~(c\overline{c}) ^{1}D_2^{(  8)}+s,d\to~^{1}D_2+X$&$2$&$2$&$0$&$v^4$\\
$b\to~(c\overline{c}) ^{3}D_J^{(  8)}+s,d\to~^{1}D_2+X$&$2$&$0$&$1$&$v^4$\\
$b\to~(c\overline{c}) ^{3}P_J^{(1,8)}+s,d\to~^{1}D_2+X$&$1$&$1$&$1$&$v^4$\\
$b\to~(c\overline{c}) ^{1}S_0^{(1,8)}+s,d\to~^{1}D_2+X$&$0$&$2$&$0$&$v^2$\\
$b\to~(c\overline{c}) ^{3}S_1^{(1,8)}+s,d\to~^{1}D_2+X$&$0$&$2$&$1$&$v^4$\\
\end{tabular}
\caption{The velocity scalings of the amplitudes 
for the the various subprocesses relevant to 
$B\to~^3D_J+X$ and  $B\to~^1D_2+X$. $L$ is  the number of 
derivatives in the NRQCD 4-quark operators ($O^{H}({^{2S+1}L_J})$), and the 
$E$ and $M$ are the numbers of the chromoelectric E1 and the chromomagnetic 
M1 transitions in order to reach the color-singlet  physical $D-$wave state.
 }
\label{table:scaling}
\end{table}

\begin{references}
\bibitem{kuhn} M.B. Wise, Phys. Lett. {\bf B89} (1980) 229; 
J.H. Kuhn and R\"{u}ckl, Phys. Lett. {\bf B135} (1984) 477.
\bibitem{NRQCD} G.T. Bodwin, E. Braaten and G.P. Lepage, Phys. Rev.
{\bf D51} (1995) 1125.
\bibitem{pdecay} G.T. Bodwin, E. Braaten and G.P. Lepage, Phys. Rev. 
{\bf D46} (1992) 1914.
\bibitem{bodwinB}  G.T. Bodwin, E. Braaten, T.C. Yuan and P. Lepage,
Phys. Rev. {\bf D46} (1992) R3703.
\bibitem{jungilb} Pyungwon Ko, Jungil Lee and H. S. Song,
Phys. Rev. {\bf D53} (1996) 1409.
\bibitem{flemingB} Sean Fleming, Oscar F. Hernandez, Ivan Maksymyk
and Helene Nadeau, MADPH-96-953, hep-ph/9608413. 
\bibitem{rosner} W. Kwong, J.L. Rosner and C. Quigg, Ann. Rev. Nucl. 
and Part. Sci., {\bf 37} (1987) 325.  
\bibitem{ko2p} Pyungwon Ko, Phys. Rev. {\bf D52} (1995) 3108; 
               S. F. Tuan,  Pramana J. Phys. {\bf 45} (1995) 209.
\bibitem{qiao} C.-F. Qiao, F. Yuan and K.-T. Chao, PUTP-96-28, hep-ph/9609284.
\bibitem{pcho2}P. Cho, A. K. Leibovich, Phys. Rev. {\bf D53} (1996) 6203.
\bibitem{wise} P. Cho and M.B. Wise, Phys. Rev. {\bf D51}  (1995) 3352.  
\end{references}
\end{document}